\documentclass{aa}
\usepackage{graphics}
\usepackage{psfig}

\def\hh{H$_2$}
\def\co{C$^{17}$O}
\def\cs{C$^{34}$S}
\def\water{H$_2$O}
\def\hcop{HCO$^+$}
\def\hcopp{H$^{13}$CO$^+$}

\def\hhh{H$_3^+$}

\def\zet{$\zeta_{\rm CR}$}
\def\op{$o$/$p$}

\def\gtsim{{_>\atop{^\sim}}}
\def\ltsim{{_<\atop{^\sim}}}

\def\kms{km~s$^{-1}$}

\def\scm{cm$^{-2}$}
\def\ccm{cm$^{-3}$}

\def\mic{$\mu$m}

\def\msol{M$_{\odot}$}
\def\lsol{L$_{\odot}$}

\newcommand\araa{{ARA\&A}}
\newcommand\aap{{A\&A}}

\newcommand\aaps{{A\&AS}}

\newcommand\apj{{ApJ}}
\newcommand\apjl{{ApJ}}
\newcommand\apjs{{ApJS}}

\newcommand\nat{{Nature}}

\begin{document}

\thesaurus{08.06.2; 08.03.4; 09.03.2; 09.13.2; 09.19.1}

\title{Limits on the cosmic-ray ionization rate toward massive young stars}
\titlerunning{The cosmic-ray ionization rate toward massive protostars}
\author{Floris F.S. van der Tak \and Ewine F. van Dishoeck}
\institute{Sterrewacht, Postbus 9513, 2300 RA Leiden, The Netherlands}
\authorrunning{van der Tak \& van Dishoeck}
\offprints{Floris van der Tak \\ (vdtak@strw.leidenuniv.nl)}
\date{Received May 5, 2000 / Accepted May 27, 2000}

\maketitle

\begin{abstract}
  
  Recent models of the envelopes of seven massive protostars are used
  to analyze observations of H$_3^+$ infrared absorption and \hcopp\ 
  submillimeter emission lines toward these stars, and to constrain
  the cosmic-ray ionization rate \zet. The \hcopp\ gives best-fit
  values of \zet$=(2.6 \pm 1.8) \times 10^{-17}$~s$^{-1}$, in good
  agreement with diffuse cloud models and with recent Voyager/Pioneer
  data but factors of up to 7 lower than found from the H$_3^+$ data.
  No relation of \zet\ with luminosity or total column density is
  found, so that local (X-ray) ionization and shielding against cosmic
  rays appear unimportant for these sources.  The difference between
  the \hhh\ and \hcopp\ results and the correlation of $N$(\hhh) with
  heliocentric distance suggest that intervening clouds contribute
  significantly to the \hhh\ absorptions in the more distant regions.
  The most likely absorbers are low-density ($\ltsim 10^4$~\ccm)
  clouds with most carbon in neutral form or in CO.
  

\end{abstract}

\section{Introduction}
\label{s:intro}

The ionization fraction of molecular clouds is an important parameter
for their dynamics through its control over the influence of any
magnetic field.  The ionization also has a major effect on the
chemistry of molecular clouds because ion-neutral reactions are
generally much faster than neutral-neutral reactions. In dense regions
shielded from direct ultraviolet irradiation, the ionization is
dominated by cosmic rays. However, the rate of this process \zet\ has
not yet been constrained directly. The current best estimate comes
from chemical models to reproduce the observed abundances of OH and HD
in diffuse interstellar clouds (Hartquist et al.\ 1978; van Dishoeck
\& Black 1986; Federman et al.\ 1996)\nocite{hart78,evd86,feder96},
notably those toward Perseus OB2.  These models indicate that
\zet$=10^{-16} - 10^{-17}$~s$^{-1}$ per H~atom, but with a factor
of~10 uncertainty because of uncertainties in temperature, radiation
field and the effects of shocks.  In addition, it is unknown if \zet\
varies with location in the Galaxy, since the diffuse cloud results
are limited to the solar neighbourhood.

Each cosmic-ray ionization of H$_2$ yields one \hhh\ molecule, so that
\hhh\ has a constant concentration which depends only on \zet\ and the
abundances of its main destroyers: CO, O and electrons. Hence, the
recent detections of \hhh\ infrared absorption lines by \cite{gebal96}
and \cite{bmcc99} toward massive protostars provide a novel way to
measure \zet. We constrain \zet\ using models by
\cite{fvdt00} of the envelopes of these stars, and use the
derived values to model observations of \hcop, the abundance of which
is also proportional to \zet.

\begin{figure}[b]
  \begin{center}

\psfig{file=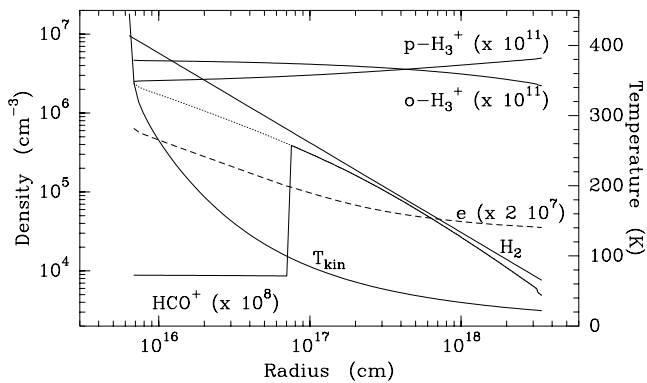,width=9cm,angle=-90}
    
    \caption{Temperature and density structure of GL 2136, and
      calculated concentrations of \hhh, $e$ and \hcop, both with
      (full line) and without (dotted line) destruction by \water\ 
      included.}
    \label{fig:gl2136}
  \end{center}
\end{figure}


\begin{table*}[!pt]
\caption{Observed and modeled column densities and ortho/para ratios
  of \hhh\ and CO.}
\begin{center}
    \begin{tabular}{llccccccc}
\hline
Source & \multicolumn{3}{c}{Observed$^{(a)}$} & \multicolumn{3}{c}{Modeled$^{(b)}$} 
                                                           & \multicolumn{2}{c}{Inferred \zet$^{(c)}$} \\ 
 & \multicolumn{3}{c}{\hrulefill}  & \multicolumn{3}{c}{\hrulefill} & \multicolumn{2}{c}{\hrulefill} \\ 
 & $N$(\hhh)  & \op & $N$(CO) & $N$(\hhh)  & \op & $N$(CO) & (1) & (2) \\
 & $10^{13}$~\scm & ratio & $10^{19}$~\scm & $10^{13}$~\scm& ratio & $10^{19}$~\scm 
                                                           & \multicolumn{2}{c}{$10^{-17}$ s$^{-1}$} \\ \hline

GL 2136  &$ 38\pm 4$  & $1.0^{+0.4}_{-0.3}$ & 2.2 & $3.9$ & $0.45$ & 3.6 & $9.7$  & $ 16$ \\
GL 2591  &$ 22\pm 2$  & $0.8^{+0.2}_{-0.1}$ & 1.3 & $2.0$ & $0.46$ & 2.1 & $11$ & $ 18$ \\
GL 490   &$ 11\pm 6$  & $0.6^{+1.6}_{-0.5}$ & 0.78 & $5.2$ & $0.19$ &8.4 & $2.1$  & $ 23$ \\
W 33A    &$ 52\pm 13$ & $0.8^{+0.5}_{-0.3}$ & 2.6 &$16.6$ & $0.26$ & 7.8 & $3.1$  & $ 9.3$ \\
NGC 2264 &$<12$  & --                       & 2.2 & $3.1$ & $0.29$ & 7.2 & $<3.9$ & $<13$ \\
W3 IRS5  &$<8.6$ & --                       & 2.6 & $4.3$ & $0.51$ & 9.4 & $<2.0$ & $<7.2$ \\
S 140    &$<4.4$ & --                       & 0.74& $2.3$ & $0.41$ & 6.3 & $<1.9$ & $<16$ \\ \hline

    \end{tabular}
    \label{t:res}

\end{center}

{\scriptsize $^a$} \hhh\ from \cite{bmcc99}; CO from 
$^{13}$CO observations by Mitchell
et al.\ (1990, 1993, 1995)\nocite{mitc90,mitc93,mitc95}.

{\scriptsize $^b$} Using the
physical structure from \cite{fvdt00} and assuming \zet\ $=1\times 10^{-17}$~s$^{-1}$.

{\scriptsize $^c$} Case (1): $(N_{\rm obs}$(\hhh) $/ N_{\rm model}$(\hhh))  $\times 10^{-17}$ s$^{-1}$;
           case (2) = case (1) $\times (N_{\rm obs}$(CO) $/ N_{\rm model}$(CO)).

\end{table*}

\section{Models}
\label{s:model}

McCall et al.\ (1999)\nocite{bmcc99} present observations of
rovibrational lines of \hhh\ in absorption against seven luminous
($10^4-10^5$~\lsol) young stars, which are still embedded in envelopes
of $\sim 100$~\msol\ of dust and molecular gas. These same sources
have been studied by \cite{mitc90} in $^{13}$CO infrared absorption
and by \cite{fvdt00} in submillimeter dust continuum and CS, \cs\ and
\co\ line emission. Based on these data sets, van der Tak et al.\ 
(1999, 2000)\nocite{fvdt99,fvdt00} modeled the temperature and density
structure of the envelopes using a power law structure $n=n_0
(r/r_0)^{-\alpha}$. The radial dust temperature profile is calculated
self-consistently from the luminosity and $n_0$ is determined from
submillimeter photometry which probes the dust column density. The
parameter $\alpha$ is constrained by modeling the relative strengths
of the CS and C$^{34}$S $J=2\to 1$ through $10\to 9$ lines with a
non-LTE radiative transfer program based on the Monte Carlo method. 

The outer radii of the models are twice the half-intensity radii of
the CS $J=5\to4$ emission, given in Table~5 of \cite{fvdt00}.
For the sources studied here, the values are $(3-11)\times 10^{17}$~cm,
which are accurate to a factor of 2. However, the dust and gas maps
also reveal emission extending outside the envelopes of W~3 IRS5 and
GL~490, and maybe W~33A. No ``skin'' appears to surround GL~2591 and
GL~2136. The case of S~140 is complicated: the dust appears to be
heated by multiple sources, and its emission is not well fitted by a
centrally heated model. We do not have a dust map of NGC~2264, but
this source is part of an extended molecular cloud complex, so that
extended material may also contribute to the \hhh\ absorption.
Evidence for extended components at lower temperature and/or column
density than those probed by the dust emission comes from
emission in low-$J$ lines of CO at $\gtsim 1'$ offsets, and
self-absorptions on their central line profiles. These features are
present in the data for all the sources discussed in this paper.
Before considering foreground contributions in \S~\ref{s:skin}, we
concentrate on the dense molecular envelopes.

\section{Results}
\label{s:res}

Given the temperature and density profiles, we calculate the \hhh\ 
concentration at each position in the envelopes.  Considering only
cosmic rays as producers of \hhh\ and reactions with CO and O as
destroyers, the concentration of \hhh\ is given by $n$(\hhh)=
\zet/[$x$(CO)$\times k_{\rm CO}$ + $x$(O)$\times k_{\rm O}$].  In this
expression, $x$(CO) and $x$(O) are the abundances of CO and O relative
to \hh, and $k_{\rm CO}$ and $k_{\rm O}$ the rate coefficients for
their respective reactions with \hhh, taken from \cite{mill97}. We
neglect any dependence of $k_{\rm CO}$ on temperature since the dipole
moment of CO is small.  The models use an abundance of CO of $2\times
10^{-4}$ at temperatures above $20$~K, and zero below due to
freeze-out on dust grains.  This abundance behaviour is consistent
with observations of \co\ emission lines by \cite{fvdt00}.  The
abundance of O is assumed to be $1.5\times 10^{-4}$ based on the
models of \cite{lee96}, and the temperature in our models does not
drop below 14~K, where O would freeze out.  The ortho/para (\op) or
$(J,K)=(1,0)/(1,1)$ ratio of \hhh\ changes with radius since the
ground state of ortho-\hhh\ lies $32.86$~K above that of para-\hhh\ 
(\cite{dinel97}), and reactive collisions with \hh\ tie the \op\ 
ratio to the kinetic temperature.  Figure~\ref{fig:gl2136} illustrates
the results for the case of GL~2136.

Integration of the concentrations of ortho- and para-\hhh\ over radius
yields total column densities $N$(\hhh) and mean \op\ ratios, which
are compared with the data in Table~\ref{t:res}. The calculated \op\ 
ratios are consistent with the data within the observational errors,
but the model values are systematically lower than those observed.
For CO, the models, which were constrained by emission data, typically
overproduce absorption measurements of $N$(CO) by factors of~3,
probably due to deviations from spherical geometry on small scales,
consistent with several other tracers (van der Tak et al.\ 
2000)\nocite{fvdt00}. Since the model values of $N$(\hhh) may be less
affected because \hhh\ is more evenly distributed than CO (Fig.~1),
Table~\ref{t:res} presents the values of \zet\ both before (case 1)
and after (case 2) scaling the model down by the ratio of observed to
modeled $N$(CO).  The uncertainty in the model is a factor of two due
to the uncertain radii of the envelopes.  The estimates of \zet\ in
case (2) are considerably larger than those in previous work
(\S~\ref{s:intro}), which together with the low \op\ ratios indicates
that there may be an additional component of warm \hhh\ along the line
of sight.  We will estimate the contribution to $N$(\hhh) by the dense
envelopes by modeling emission lines of \hcopp\ which have critical
densities of $\sim 10^6$~\ccm, and hence cannot arise in the
foreground.

\begin{figure}[b]

\psfig{figure=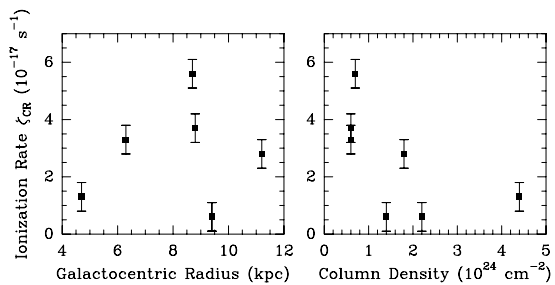,width=8cm,angle=-90}
    
    \caption{Derived cosmic-ray ionization rates versus
      galactocentic radius \textit{(left)} and $N$(\hh) in a 
   $15''$ beam \textit{(right)}.}
    \label{fig:corr}
\end{figure}

\section{Comparison with HCO$^+$}
\label{s:hcop}

In the dense envelopes, the main destruction route of \hhh\ is the
reaction with CO into \hcop. The concentration of \hcop\ is given by
$n$(\hcop)=$x$(CO)$n$(\hhh)$k_{\rm CO}$/ [$(x(e)k_{e}$ +
$x$(\water)$k_{{\rm H}_2{\rm O}}$], with $k_e$ the rate coefficient
for dissociative recombination of \hcop. The electron fraction $x(e)$
has been calculated at each point in the envelopes with a small
chemical network (cf.\ de Boisanger et al.\ 1996\nocite{bois96}) based
on the UMIST reaction rates (Millar et al.\ 1997)\nocite{mill97}. The
main difference with the analysis of de Boisanger et al.\ (1996) is
the use of a detailed physical structure to interpret the
high-excitation lines.  We assume that O$_2$ and \water\ have
negligible ($\ltsim 10^{-6}$) abundances in the bulk of the envelopes,
but that at $T>100$~K, $x$(\water) jumps to $5\times 10^{-5}$ due to
grain mantle evaporation. We neglect metals such as Mg, Fe and S as
contributors to $x(e)$ and large molecules such as polycyclic aromatic
hydrocarbons as sinks of $x(e)$; using the low metal abundances
inferred from dark cloud chemistry models would increase $x(e)$ by a
factor of 2--3 (Lee et al.\ 1996).  The values of \zet\ derived above
give $x(e) \sim 10^{-7}$ at the outer radii and $\sim 10^{-9}$ at the
inner radii, as illustrated in Fig.~\ref{fig:gl2136}.  The precipitous
drop of \hcop\ at $100$~K, caused by reactions with evaporated
water, occurs at too small radii to affect our results.

\begin{table}[t]

\caption{Observed and modeled fluxes $\int T_{\rm mb}dV$
  (K~\kms) of the \hcopp\ $J=3\to2$ and $4\to3$ lines.}

    \begin{tabular}{llccccc}
\hline
Source & \multicolumn{2}{c}{Observed} & \multicolumn{2}{c}{Modeled} & \zet & $N$(\hhh) \\
       & $3-2$  & $4-3$  & $3-2$  & $4-3$ & {\scriptsize (a)} & {\scriptsize (b)} \\ \hline

GL 2136  & 4.1 & 4.0 &  4.5 &  2.9 & 3.3 &  8 \\
GL 2591  & 5.5 & 4.4 &  5.5 &  3.2 & 5.6 &  7 \\
GL 490   & 2.4 & 1.9 &  3.0 &  2.0 & 0.64 & 0.3 \\
W 33A    & 8.9 & 5.3 &  9.8 &  5.3 & 1.3 &  7 \\
NGC 2264 & 4.0 & 2.1 &  3.3 &  2.4 & 0.61 & 0.6 \\
W3 IRS5  & 9.3 & 9.2 & 8.2  &  6.0 & 2.8 &  3 \\
S 140    & 9.2 & 8.6 &  8.5 &  6.6 & 3.7 &  1.1 \\
\hline
    \end{tabular}

{\scriptsize $(a)$:} Best fit value of \zet\ to \hcopp\ data, in $10^{-17}$ s$^{-1}$.

{\scriptsize $(b)$:} Predicted from \hcopp\ for case (2), in $10^{13}$~\scm.

    \label{t:hco+}
\end{table}

In the comparison with data, we use the $60\times$ less abundant
isotope \hcopp\ to avoid optical depth effects. The maximum optical
depth in the lines is $\approx 1$ in our models.  Table~\ref{t:hco+}
lists the calculated fluxes of the \hcopp\ $J$=3$\to$2 and 4$\to$3
lines in $18''$ and $14''$ beams.  Observations are from \cite{fvdt99}
for GL~2591 and from \cite{bois96} for NGC~2264 and W~3 IRS5. The data
for W~33A, GL~490, S~140 and GL~2136 were obtained with the James
Clerk Maxwell Telescope in the way described in \cite{fvdt00}.  Using
\zet\ derived from \hhh, the models overproduce \hcop\ by factors of
$2-7$. Adjusting the models to the \hcopp\ data yields refined
estimates for \zet\ (Table~\ref{t:hco+}) which pertain strictly to the
dense molecular gas, unaffected by any intervening clouds along the
line of sight. The data for the various sources span the range of
\zet$=(2.6 \pm 1.8) \times 10^{-17}$~s$^{-1}$, in good agreement with
the diffuse cloud estimates (\S~\ref{s:intro}), and also consistent
with recent data from the Voyager and Pioneer spacecraft at distances
up to $60$~AU from the Sun (\cite{webb98}).

Figure~\ref{fig:corr} shows that the source-to-source variation in
\zet\ is not related to Galactic structure through differences in
cosmic-ray flux, nor to shielding against cosmic rays at high \hh\ 
column densities. The values of \zet\ are also unrelated to
luminosity, which implies that local ionization such as by X-rays
(\cite{phil96}) is unimportant on the scales traced by our data.
Variations of the cosmic-ray density by $50$\% on scales of a few kpc
are in good agreement with results from $\gamma-$ray observations
(e.g., Hunter et al.\ 1997).\nocite{hunt97} However, why does \hhh\
give systematically higher values of \zet?

\begin{figure}[t]
  \begin{center}

\psfig{figure=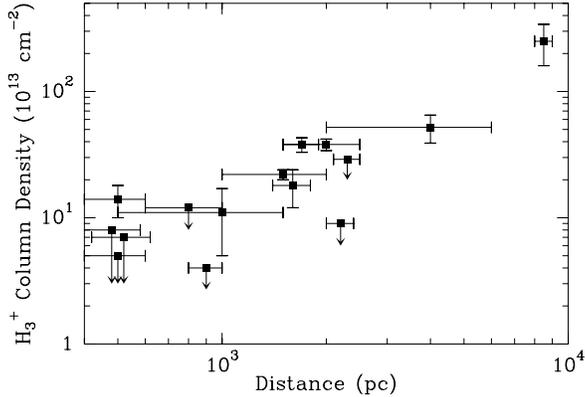,width=8cm,angle=-90}
    
    \caption{Observed $N$(\hhh) versus heliocentric distance. The
      correlation suggests that intervening clouds are important.}
   \label{fig:helio} 
\end{center}
\end{figure}

\section{Contributions by foreground layers}
\label{s:skin}

Figure~\ref{fig:helio} plots the observed $N$(\hhh) versus
heliocentric distance, including all data from \cite{bmcc99} as well
as the results for the Galactic Center and the diffuse cloud in front
of Cyg~OB2 \#12 from \cite{gebal99}.  The dense cloud data have a
correlation coefficient of 93\%, suggesting that absorption by
intervening clouds plays an important role for the more distant
sources. This section investigates the possible nature of these
absorbers.

First, the absorptions may occur at the edges of the dense cores
studied here, where carbon is in neutral or ionized form.  This
``photodissociation region'' occupies $\sim$3--4 magnitudes of visual
extinction (\cite{holl97}), corresponding to $N_H \ltsim 8 \times
10^{21}$~\scm.  The ionized layer is negligible because the high
electron fraction ($\sim 10^{-4}$) limits $n$(\hhh) to $10^{-7}$~\ccm.
For the neutral component, assuming $n\sim 10^4$~\ccm\ as derived
specifically for S~140 by \cite{timm96}, and $n$(\hhh) $\sim
10^{-4}$~\ccm, we find $N$(\hhh) $\sim$ few $\times 10^{13}$~\scm,
comparable to the dense envelopes.

Second, the absorbers may consist of cold ($\ltsim 20$~K) molecular
gas. For $n \gtsim 10^4$~\ccm, the CO will be frozen out on the
grains. \cite{tiel91} observed solid CO in absorption toward all our
sources and found $N$(CO)$\sim 10^{17}$~\scm, or $N_{\rm H}\sim
10^{21}$~\scm, assuming that most carbon is in solid CO. The implied
column lengths are too short to be of importance for \hhh, and
the low temperatures are incompatible with the observed $o$/$p$
H$_3^+$ ratios.

Third, the H$_3^+$ absorptions may arise in clouds with $n \ltsim
10^4$ cm$^{-3}$, which either surround the power-law envelopes or
happen to lie along the line of sight.  Such ``translucent''
foregrounds are visible in our data (\S~\ref{s:model}), and can
contribute $N$(\hhh) $\sim 10^{14}$~\scm\ each based on models by
\cite{vdb89}.  These tenuous clouds have long path lengths and may
dominate the H$_3^+$ absorption. At low densities, \hcop\ may form
through OH + C$^+$ $\to$ H + CO$^+$ followed by \hh\ + CO$^+ \to$ H +
\hcop.  However, for our sources, [CII] $158$~\mic\ data indicate
$N$(C$^+$)/$N$(CO) $\ltsim 10^{-2}$.  Translucent clouds are generally
weak in \hcop\ emission (\cite{gred94}).

The velocities of the \hhh\ absorptions are consistent with those of
the submillimeter emission lines of \co\ and \cs, suggesting that the
\hhh\ absorbers are in the vicinity of the infrared sources. However,
the correlation of $N$(\hhh) with distance remains after subtracting
the dense core contribution (Table~\ref{t:hco+}), suggesting a
non-local origin.  Altogether, the data indicate that the contribution
of the envelopes to $N$(\hhh) varies from $\sim 10^{13}$ to $\sim
10^{14}$~\scm, and that any additional absorption seen in sources at
$d>2$~kpc occurs in intervening clouds.


In summary, observations of \hhh\ absorption and \hcopp\ emission
lines, combined with models of the temperature and density structure
of the sources, constrain the cosmic-ray ionization rate to \zet
$=(2.6 \pm 1.8) \times 10^{-17}$~s$^{-1}$, with upper limits that are
factors of 3--5 higher.  Future tests of the results include more
sensitive observations of \hhh\ toward W~3 IRS5 and NGC 2264,
velocity-resolved observations to search for \hhh\ absorption at
offset velocities from the dense cores, observations of more distant
sources to test the correlation with distance, and observations of
\hcop\ infrared absorption lines to directly compare with H$_3^+$ and
CO infrared absorption.

\begin{acknowledgements} 
  
  We thank Neal Evans, Dan Jaffe, Tom Geballe and John Black for useful
  discussions.  This research is supported by NWO grant 614-41-003.
  
\end{acknowledgements}

\end{document}